\documentclass{fizik}
\usepackage{multicol,epsfig}

\vol{XX}
\pyear{XXXX}
\received{July 2003}

\author[VAN VELSEN, KINDERMANN, BEENAKKER]{
\textbf{J.L. van Velsen, M. Kindermann, and C.W.J. Beenakker}\\
\textit{Instituut-Lorentz, Universiteit Leiden,}\\
\textit{P.O. Box 9506, 2300 RA Leiden, THE NETHERLANDS}
}

\title{Dephasing of entangled electron-hole pairs in a degenerate electron gas}

\begin{document}
\maketitle

\begin{abstract}
A tunnel barrier in a degenerate electron gas was recently discovered as a source of entangled electron-hole pairs. Here, we
investigate the loss of entanglement by dephasing. We calculate both the maximal violation ${\cal E}_{\rm max}$ of the Bell inequality
and the degree of entanglement (concurrence) ${\cal C}$. 
If the initially maximally entangled electron-hole pair is in a Bell state, then the Bell inequality is violated for arbitrary strong
dephasing. The same relation ${\cal E}_{\rm max}=2\sqrt{1+{\cal C}^{2}}$ then holds as in the absence of dephasing. More generally, for
a maximally entangled superposition of Bell states, the Bell inequality is satisfied for a finite dephasing strength and 
the entanglement vanishes for somewhat stronger
(but still finite) dephasing strength. There is then no one-to-one relation between ${\cal E}_{\rm max}$ and ${\cal C}$.

\keywords{Entanglement, Bell inequality, Nonlocality, Decoherence}
\end{abstract}

\section{Introduction}

The production and detection of entangled particles is the essence of quantum information processing \cite{Nie00}.
In optics, this is well-established with polarization-entangled photon pairs, but in the solid state it remains an experimental challenge. 
There exist several theoretical proposals for the production and detection of entangled electrons \cite{Egu03,Mar03}. 
These theoretical works address mainly pure states.
The purpose of this article is to investigate what happens if the state is mixed. 
Some aspects of this problem were also considered in Refs.\ \cite{Bur03,Sam03,Bee03}.
We go a bit further by comparing violation of the Bell inequality to the degree of entanglement of the mixed state.
 
The Bell inequality is a test for the existence of nonclassical correlations in a state shared by two spatially separated observers \cite{Bel64}. It is called an entanglement ``witness'', because violation of the inequality implies that the state is quantum mechanically entangled --- but not the other way around \cite{Ter03}. More precisely, while all entangled pure states violate the Bell inequality, there exist mixed states which are entangled and nevertheless satisfy the inequality \cite{Wer89}. A mixed state can arise either because of the interaction with an environment (proper mixture) or because the detector does not differentiate among certain degrees of freedom of the entangled pure state (improper mixture). Generically, the loss of purity of a state is associated with a decrease in the degree of entanglement (although this is not necessarily so).

Applications of these general notions typically involve polarization-entangled photon pairs \cite{Man95}. The transition from pure to mixed states, and the associated degradation of entanglement, can be avoided quite effectively in that context --- even if the photons interact strongly with matter degrees of freedom. For a dramatic demonstration, see a recent experiment \cite{Alt02} and theory \cite{Vel02} on plasmon-assisted entanglement transfer. In essence, this robustness of photon entanglement is a manifestation of the fact that linear optics is an excellent approximation even if the medium in which the photons propagate is strongly scattering and absorbing.

The entanglement scheme that we will 
analyze here, proposed in Ref.\ \cite{Bee03}, 
involves the Landau level index of an electron and hole quasiparticle. The scheme differs from earlier 
proposals in that the entanglement is produced by a single-electron Hamiltonian, without requiring Coulomb interaction or the superconductor 
pairing interaction. 
We consider one specific mechanism for the loss of purity, namely interaction with the environment. We model this interaction 
phenomenologically by introducing phase factors in the scattering matrix and subsequently averaging over these phases. 
A more microscopic treatment (for example along the lines of a recent paper \cite{Marq03}) is not attempted here.
The mixed state created by this averaging is a proper mixture. 
An improper mixture would result from energy averaging.
We assume that the applied voltage is sufficiently small that we can neglect energy averaging. Experimentally, both energy and phase 
averaging may play a role \cite{Ji03}.

\section{Description of the edge state entangler}

\begin{figure}[htb]
\begin{center}
\includegraphics[width=0.5\linewidth]{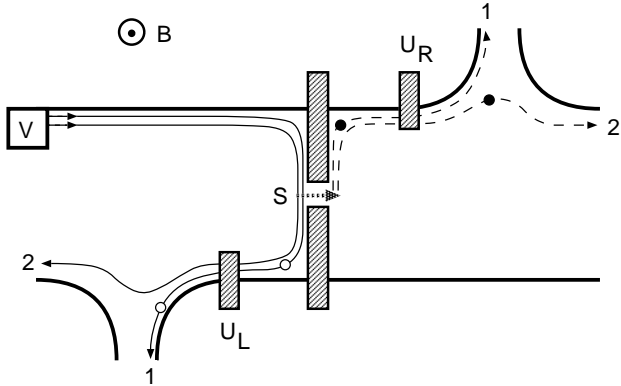}
\end{center}
\caption{Schematic drawing of the edge state entangler. Taken from Ref.\ \protect\cite{Bee03}.}
\label{bell_circuit}
\end{figure}

In Fig.\ \ref{bell_circuit} we illustrate the method to produce and detect entangled edge states in the quantum Hall effect \cite{Bee03}. The thick black lines indicate the boundaries of a two-dimensional electron gas. A strong perpendicular magnetic field $B$ ensures that the transport near the Fermi level $E_{F}$ takes place in two edge channels, extended along a pair of equipotentials (thin solid and dashed lines, with arrows that give the direction of propagation). A split gate electrode (shaded rectangles at the center) divides the conductor into two halves, coupled by tunneling through a narrow opening (dashed arrow, scattering matrix $S$). If a voltage $V$ is applied between the two halves, then there is a narrow energy range $0<\varepsilon<eV$ above $E_{F}$ in which the edge channels are predominantly filled in the left half (solid lines) and predominantly empty in the right half (dashed lines). 

Tunneling events introduce filled states in the right half [black dots, creation operator $b^{\dagger}_{i}(\varepsilon)$] and empty states in the left half [open circles, creation operator $c^{\dagger}_{i}(\varepsilon)$]. These are quasiparticle excitations of the vacuum state $|0\rangle_{\varepsilon}$, corresponding to empty states in the left half and filled states in the right half. To leading order in the tunneling probability the wavefunction is given by
\begin{eqnarray}
&&|\Psi\rangle=\prod_{\varepsilon}\bigl(\sqrt{w}|\Phi\rangle_{\varepsilon}+\sqrt{1-w}|0\rangle_{\varepsilon}\bigr),\label{Psidef}\\
&&|\Phi\rangle_{\varepsilon}=w^{-1/2}\sum_{i,j}c_{i}^{\dagger}(\varepsilon)\gamma_{ij}b_{j}^{\dagger}(\varepsilon)|0\rangle_{\varepsilon},\label{Phidef}\\
&&
\gamma=\sigma_{y} r\sigma_{y}t^{\rm T},\;\;w={\rm Tr}\,\gamma\gamma^{\dagger}. \label{gammadef}
\end{eqnarray}
The matrix $\gamma$ is given in terms of a Pauli matrix, 
\begin{equation}
\sigma_{x}=\left(\begin{array}{cc}
0&1\\1&0
\end{array}\right) \equiv \sigma_{1},\;\;
\sigma_{y}=\left(\begin{array}{cc}
0&-i\\i&0
\end{array}\right) \equiv \sigma_{2},\;\;
\sigma_{z}=\left(\begin{array}{cc}
1&0\\0&-1
\end{array}\right) \equiv \sigma_{3} \label{sigmadef},
\end{equation}
and the reflection and transmission matrices $r,t$. (These are $2\times 2$ submatrices of $S$.) The state $|\Psi\rangle$ is a superposition of the vacuum state $|0\rangle$ and the entangled particle-hole state $|\Phi\rangle$. Terms containing two particles or two holes are of higher order in the tunneling probability and can be neglected. 
We also assume that the applied voltage is sufficiently small that the energy dependence of the scattering matrix need not be taken into
account.

Dephasing is introduced phenomenologically through random phase shifts $\phi_{i}$ ($\psi_{i}$) accumulated in channel $i$ at the
left (right) of the tunnel barrier. The reflection and transmission matrices transform as
\begin{equation}
r\rightarrow\left(\begin{array}{cc} 
e^{i\phi_{1}} & 0 \\
0 & e^{i\phi_{2}} 
\end{array}\right)r_{0},\;\;t\rightarrow\left(\begin{array}{cc} 
e^{i\psi_{1}} & 0 \\
0 & e^{i\psi_{2}} 
\end{array}\right)t_{0}.\label{rtepsilon}
\end{equation}
By averaging over the phase shifts, with distribution $P(\phi_{1},\phi_{2},\psi_{1},\psi_{2})$, the pure state (\ref{Psidef})  
is converted into a mixed state. 
Projecting out the vacuum contribution (which does not contribute to current fluctuations), we obtain for this
mixed state the $4 \times 4$ density matrix
\begin{equation}
\rho_{ij,kl}=\frac{\langle \gamma_{ij}^{\vphantom{\ast}}\gamma^{\ast}_{kl}\rangle}{\langle {\rm Tr}\,\gamma\gamma^{\dagger} \rangle},
\label{rhodef}
\end{equation}
where $\langle \cdots \rangle$ denotes the average over the phases.
The degree of entanglement is quantified by the concurrence ${\cal C}$, given by \cite{Woo98}
\begin{equation}
{\cal C}=\rm{max}\left\{0,\sqrt{\lambda_{1}}-\sqrt{\lambda_{2}}-\sqrt{\lambda_{3}}-\sqrt{\lambda_{4}}\right\}. \label{Cdef}
\end{equation}
The $\lambda_{i}$'s are the eigenvalues of the matrix product $\rho\cdot (\sigma_{y}\otimes\sigma_{y})\cdot\rho^{*}\cdot 
(\sigma_{y}\otimes\sigma_{y})$, in the order $\lambda_{1}\geq\lambda_{2}\geq\lambda_{3}\geq\lambda_{4}$.
The concurrence ranges from 0 (no entanglement) to 1 (maximal entanglement).

The entanglement of the particle-hole excitations is detected by the violation of the Bell-CHSH (Clauser-Horne-Shimony-Holt) inequality \cite{Cla69,Cht02}. This requires two gate electrodes to locally mix the edge channels (scattering matrices $U_{L}$, $U_{R}$) and two pairs of contacts $1,2$ to separately measure the current fluctuations $\delta I_{L,i}$ and $\delta I_{R,i}$ ($i=1,2$) in each transmitted and reflected edge channel. In the tunneling regime the Bell inequality can be formulated in terms of the low-frequency noise correlator \cite{Sam03}
\begin{equation}
C_{ij}=\int_{-\infty}^{\infty}dt\,\overline{\delta I_{L,i}(t)\delta I_{R,j}(0)}.\label{Cijdef}
\end{equation}
At low temperatures ($kT\ll eV$) the correlator has the general expression \cite{But90}
\begin{equation}
C_{ij}(U_{L},U_{R})=-(e^{3}V/h)\, \left|\left(U_{L}rt^{\dagger}U_{R}^{\dagger}\right)_{ij}\right|^{2}. \label{CijUdef}
\end{equation}
We again introduce the random phase shifts into $r$ and $t$ and average the correlator.
The Bell-CHSH parameter is
\begin{equation}
{\cal E}=|E(U_{L},U_{R})+E(U'_{L},U_{R})+E(U_{L},U'_{R})-E(U'_{L},U'_{R})|,\label{CHSHdef}
\end{equation}
where $E(U,V)$ is related to the average correlators $\langle C_{ij}(U,V) \rangle$ by
\begin{equation}
E=\frac{\langle  C_{11}+C_{22}-C_{12}-C_{21} \rangle}{\langle C_{11}+C_{22}+C_{12}+C_{21} \rangle}.\label{Edef}
\end{equation}
The state is entangled if ${\cal E}>2$ for some set of $2\times 2$ unitary matrices $U_{L}, U_{R}, U'_{L}, U'_{R}$. If ${\cal E}=
2\sqrt{2}$ the entanglement is maximal.

\section{Calculation of the mixed-state entanglement}
We simplify the problem by assuming that the two transmission eigenvalues (eigenvalues of $tt^{\dagger}$) are 
identical: $T_{1}=T_{2}\equiv T$. In the absence of dephasing the electron and hole then form a maximally entangled pair. The transmission
matrix $t_{0}=T^{1/2}V$ and reflection matrix $r_{0}=(1-T)^{1/2}V'$ in this case are equal to a scalar times a unitary matrix $V,V'$. 
Any $2\times 2$ unitary matrix $\Omega$ can be parameterized by
\begin{equation}
\Omega=e^{i\theta}
\left(\begin{array}{cc}
e^{i\alpha}&0\\
0&e^{-i\alpha}
\end{array}\right)
\left(\begin{array}{cc}
\cos\xi&\sin\xi\\
-\sin\xi&\cos\xi
\end{array}\right)
\left(\begin{array}{cc}
e^{i\beta}&0\\
0&e^{-i\beta}
\end{array}\right),\label{Omegadef}
\end{equation}
in terms of four real parameters $\alpha,\beta,\theta,\xi$. The angle $\xi$ governs the extent to which $\Omega$ mixes the degrees of freedom (no mixing for $\xi=0,\pi/2$, complete mixing for $\xi=\pi/4$). 

If we set $\Omega=\sigma_{y}V'\sigma_{y}V^{\rm T}$ we obtain for the matrix $\gamma$ of Eq.\ (\ref{gammadef}) the parametrization
\begin{equation}
\gamma=e^{i\theta}\sqrt{T(1-T)}
{\renewcommand{\arraystretch}{0.3}
\left(\begin{array}{cc} 
e^{i\phi_{2}+i\alpha} & 0 \\
0 & e^{i\phi_{1}-i\alpha} 
\end{array}\right)}
\left(\begin{array}{cc}
\cos\xi&\sin\xi\\
-\sin\xi&\cos\xi
\end{array}\right)
\left(\begin{array}{cc} 
e^{i\psi_{1}+i\beta} & 0 \\
0 & e^{i\psi_{2}-i\beta} 
\end{array}\right).\label{gammaparam}
\end{equation}
In the same parametrization, the matrix $rt^{\dagger}$ which appears in Eq.\ (\ref{CijUdef}) takes the form
\begin{equation}
rt^{\dagger}=e^{i\theta'-i\theta}\sqrt{T(1-T)}
\left(\begin{array}{cc} 
e^{i\phi_{1}-i\alpha} & 0 \\
0 & e^{i\phi_{2}+i\alpha} 
\end{array}\right)
\left(\begin{array}{cc}
\cos\xi&\sin\xi\\
-\sin\xi&\cos\xi
\end{array}\right)
\left(\begin{array}{cc} 
e^{-i\psi_{1}-i\beta} & 0 \\
0 & e^{-i\psi_{2}+i\beta} 
\end{array}\right),\label{rtdaggerparam}
\end{equation}
with $e^{i\theta'}={\rm Det}\,V'$. We have used the identity
$V'V^{\dagger}=({\rm Det}\,V')(\sigma_{y}V'\sigma_{y}V^{\rm T})^{\ast}$ to relate the parametrization of $rt^{\dagger}$ to that of $\gamma$. Note that
\begin{equation}
{\rm Tr}\,\gamma\gamma^{\dagger}=2T(1-T)={\rm Tr}\,rt^{\dagger}tr^{\dagger},\label{tracegamma}
\end{equation}
independent of the phase shifts $\phi_{i}$ and $\psi_{i}$.

To average the phase factors we assume that the phase shifts at the left and the right of the tunnel barrier are independent, so
$P(\phi_{1},\phi_{2},\psi_{1},\psi_{2})=P_{L}(\phi_{1},\phi_{2})P_{R}(\psi_{1},\psi_{2})$. The complex dephasing parameters $\eta_{L}$ and
$\eta_{R}$ are defined by
\begin{equation}
\eta_{L}=\int d\phi_{1} \int d\phi_{2}\, P_{L}(\phi_{1},\phi_{2})e^{i\phi_{1}-i\phi_{2}}, \quad
\eta_{R}=\int d\psi_{1} \int d\psi_{2}\, P_{R}(\psi_{1},\psi_{2})e^{i\psi_{1}-i\psi_{2}}.
\end{equation}

The density matrix (\ref{rhodef}) of the mixed particle-hole state has, in the parametrization (\ref{gammaparam}), the elements
\begin{equation}
\rho=\frac{1}{2}\left(\begin{array}{cccc}
\cos^{2}\xi&\tilde{\eta}_{R}\cos\xi\sin\xi& -\tilde{\eta}^{\ast}_{L}\cos\xi\sin\xi& \tilde{\eta}^{\ast}_{L}\tilde{\eta}^{\vphantom{\ast}}_{R}\cos^{2}\xi\\
\tilde{\eta}^{\ast}_{R}\cos\xi\sin\xi&\sin^{2}\xi& -\tilde{\eta}^{\ast}_{L}\tilde{\eta}^{\ast}_{R}\sin^{2}\xi& \tilde{\eta}^{\ast}_{L}\cos\xi\sin\xi\\
-\tilde{\eta}_{L}\cos\xi\sin\xi& -\tilde{\eta}_{L}\tilde{\eta}_{R} \sin^{2}\xi&\sin^{2}\xi& -\tilde{\eta}_{R}\cos\xi\sin\xi\\
\tilde{\eta}^{\vphantom{\ast}}_{L}\tilde{\eta}^{\ast}_{R}\cos^{2}\xi& \tilde{\eta}_{L}\cos\xi\sin\xi&  -\tilde{\eta}^{\ast}_{R}\cos\xi\sin\xi&\cos^{2}\xi
\end{array}\right).\label{rhoresult}
\end{equation}
We have defined $\tilde{\eta}_{L}=\eta_{L}e^{-2i\alpha}$, $\tilde{\eta}_{R}=\eta_{R}e^{2i\beta}$. The concurrence ${\cal C}$, calculated from Eq.\ (\ref{Cdef}), has a complicated
expression. For $|\eta_{L}|=|\eta_{R}|\equiv \eta$ it simplifies to 
\begin{equation}
{\cal C}={\rm max}\left\{0,-\frac{1}{2}(1-\eta^{2})+\frac{1}{4}\sqrt{16\eta^{2}+2(1-\eta^{2})^{2}(1+\cos 4\xi)} \right\}.
\label{cfinal}
\end{equation}
Notice that ${\cal C}=\eta^{2}$ for $\xi=0$.

For the Bell inequality we first note that the ratio of correlators (\ref{Edef}) can be written as
\begin{equation}
E(U_{L},U_{R})=\frac{1}{2T(1-T)}\,\langle{\rm Tr}\,U_{L}^{\dagger}\sigma_{z}U_{L}^{\vphantom\dagger}rt^{\dagger}U_{
R}^{\dagger}\sigma_{z}U_{R}^{\vphantom\dagger}tr^{\dagger} \rangle.
\label{EULUR}
\end{equation}
We parameterize 
\begin{eqnarray}
U_{L}^{\dagger}\sigma_{z}U_{L}^{\vphantom\dagger}=n_{L,x}\sigma_{x}+n_{L,y}\sigma_{y}+n_{L,z}\sigma_{z}\equiv\hat{n}_{L}\cdot\vec{\sigma},\label{ULnL}\\
U_{R}^{\dagger}\sigma_{z}U_{R}^{\vphantom\dagger}=n_{R,x}\sigma_{x}+n_{R,y}\sigma_{y}+n_{R,z}\sigma_{z}\equiv\hat{n}_{R}\cdot\vec{\sigma},\label{URnR}
\end{eqnarray}
in terms of two unit vectors $\hat{n}_{L},\hat{n}_{R}$. Substituting the parametrization (\ref{rtdaggerparam}), Eq.\ (\ref{EULUR}) takes the form
\begin{equation}
E(U_{L},U_{R})=\frac{1}{2}{\rm Tr}\,
\left(\begin{array}{cc}
n_{L,z}&\tilde{\eta}_{L}^{\ast}\nu_{L}^{\ast}\\
\tilde{\eta}_{L}\nu_{L}&-n_{L,z}
\end{array}\right)
\left(\begin{array}{cc}
\cos\xi&\sin\xi\\
-\sin\xi&\cos\xi
\end{array}\right)
\left(\begin{array}{cc}
n_{R,z}&\tilde{\eta}_{R}^{\ast}\nu_{R}^{\ast}\\
\tilde{\eta}_{R}\nu_{R}&-n_{R,z}
\end{array}\right)
\left(\begin{array}{cc}
\cos\xi&-\sin\xi\\
\sin\xi&\cos\xi
\end{array}\right),
\label{EULURparam}
\end{equation}
where we have abbreviated $\nu_{L}=n_{L,x}+in_{L,y}$, $\nu_{R}=n_{R,x}+in_{R,y}$.

Comparing Eqs.\ (\ref{rhoresult}) and (\ref{EULURparam}), we see that
\begin{equation}
E(U_{L},U_{R})={\rm Tr}\,\rho \left(\hat{n}_{L}\cdot\vec{\sigma}\right)^{\rm T}\otimes\left(\hat{n}_{R}\cdot\vec{\sigma}\right).\label{EULURrho}
\end{equation}
(The transpose appears because of the transformation from electron to hole operators at the left of the barrier.)
This is an explicit demonstration that the noise correlator (\ref{Edef}) measures the density matrix (\ref{rhodef}) 
of the projected electron-hole state --- without the vacuum contribution.

The maximal value ${\cal E}_{\rm max}$ of the Bell-CHSH parameter (\ref{CHSHdef}) for an arbitrary mixed state was analyzed in Refs.\ \cite{Hor95,Ver02}. For a pure state with concurrence ${\cal C}$ one has simply ${\cal E}_{\rm max}=2\sqrt{1+{\cal C}^{2}}$ \cite{Gis91}. For a mixed state there is no one-to-one relation between ${\cal E}_{\rm max}$ and ${\cal C}$. Depending on the density matrix, ${\cal E}_{\rm max}$ can take on values between $2{\cal C}\sqrt{2}$ and $2\sqrt{1+{\cal C}^{2}}$. The general formula
\begin{equation}
{\cal E}_{\rm max}=2\sqrt{u_{1}+u_{2}}\label{Emaxformula}
\end{equation}
for the dependence of ${\cal E}_{\rm max}$ on $\rho$ involves the two largest eigenvalues $u_{1},u_{2}$ of the real symmetric 
$3\times 3$ matrix $R^{\rm T}R$ constructed from  $R_{kl}={\rm Tr}\,\rho\,\sigma_{k}\otimes\sigma_{l}$. 
For our density matrix (\ref{rhoresult}) we find from Eq.\ (\ref{Emaxformula}) a simple expression if $|\eta_{L}|=|\eta_{R}|\equiv\eta $. It
reads
\begin{equation}
{\cal E}_{\rm max}=\sqrt{2}\sqrt{(1+\eta^{2})^{2}+(1-\eta^{2})^{2}\cos 4\xi}.
\label{efinal}
\end{equation}

\section{Discussion}

The result ${\cal E}_{\rm max}=2(1+\eta^{4})^{1/2}$ which follows from Eq.\ (\ref{efinal}) for $\xi=0$ was found in Ref.\ \cite{Sam03} 
in a somewhat different context.
This corresponds to the case that the two edge channels are not mixed at the tunnel barrier. The Bell-CHSH inequality 
${\cal E}_{\rm max} \le 2$ is then violated for arbitrarily strong dephasing. 
\begin{figure}[ht!]
\begin{center}
\includegraphics[width=0.5\linewidth]{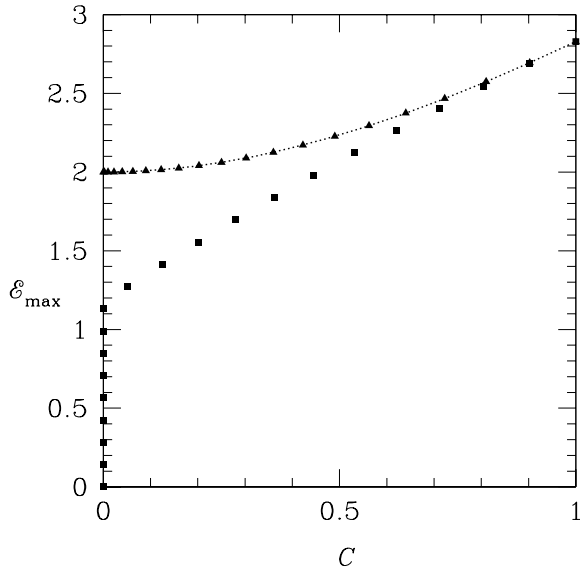}
\end{center}
\caption{
Relation between the maximal violation ${\cal E}_{\rm max}$ of the Bell-CHSH inequality and the concurrence ${\cal C}$ calculated
from Eqs.\ (\ref{cfinal}) and (\ref{efinal}) for mixing parameters $\xi=0$ (triangles, no mixing) and $\xi=\frac{\pi}{4}$ 
(squares, complete mixing). The
dephasing parameter $\eta$ decreases from 1 (upper right corner, no dephasing) to 0 (lower left, complete dephasing) with steps of
0.05. The dotted line is the relation between ${\cal E}_{\rm max}$ and ${\cal C}$ for a pure state, which is also the largest possible value
of ${\cal E}_{\rm max}$ for given ${\cal C}$.}
\label{cbell}
\end{figure}
This is not true in the more general
case $\xi \neq 0$, when ${\cal E}_{\rm max}$ drops below 2 at a finite value of $\eta$.

\begin{figure}[tbh]
\begin{center}
\includegraphics[width=0.5\linewidth]{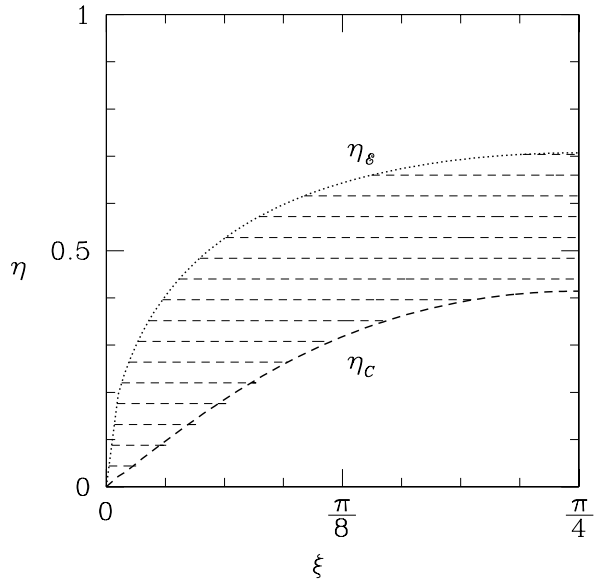}
\end{center}
\caption{The Bell-CHSH inequality is violated for dephasing parameters $\eta > \eta_{\cal E}$, while entanglement is
preserved for $\eta > \eta_{\cal C}$. The shaded region indicates dephasing and mixing parameters for which there is entanglement without
violation of the Bell-CHSH inequality.}
\label{deph}
\end{figure}

In Fig.\ \ref{cbell} we compare ${\cal E}_{\rm max}$ and ${\cal C}$ for $\xi=0$ (no mixing) and $\xi=\frac{\pi}{4}$ (complete
mixing).
For $\xi=0$ the same relation ${\cal E}_{\rm max}=2\sqrt{1+{\cal C}^{2}}$ between ${\cal E}_{\rm max}$ and ${\cal C}$ holds as for
pure states (dotted curve). Violation of the Bell inequality is then equivalent to entanglement. For $\xi \neq 0$ there exist
entangled states (${\cal C}>0$) without violation of the Bell inequality (${\cal E}_{\rm max} \le 2$). Violation of the Bell inequality
is then a sufficient but not a necessary condition for entanglement. We define two characteristic dephasing parameters 
$\eta_{\cal E}$ and $\eta_{\cal C}$ by the smallest values such that
\begin{equation}
{\cal E}_{\rm max} > 2 \quad \mbox{for} \quad \eta > \eta_{\cal E}, \quad
{\cal C} > 0 \quad \mbox{for} \quad \eta > \eta_{\cal C}.
\end{equation}
The number $\eta_{\cal E}$ is the dephasing parameter below which Bell's inequality cannot be violated; The 
dephasing parameter $\eta_{\cal C}$ gives the border between entanglement and no entanglement. 
From Eqs.\ (\ref{cfinal}) and (\ref{efinal}) we obtain
\begin{equation}
\eta_{\cal C}=\sqrt{\frac{5-\cos 4\xi -2\sqrt{2}\sqrt{3-\cos 4\xi}}{1-\cos 4\xi}}, \quad 
\eta_{\cal E}=\sqrt{\frac{-1+\cos 4\xi +\sqrt{2-2\cos 4\xi}}{1+\cos 4\xi}}.
\end{equation}
The two dephasing parameters are plotted in Fig.\ \ref{deph}. 
The inequality $\eta_{\cal E} \ge \eta_{\cal C}$ reflects the fact that ${\cal E}_{\rm max}$ is an entanglement witness. 

In conclusion, we have shown that the extent to which dephasing prevents the Bell inequality from detecting entanglement depends
on the mixing of the degrees of freedom at the tunnel barrier. No mixing ($\xi=0$) means that the maximally entangled electron-hole
pair produced by the tunnel barrier is in one of the two Bell states
\begin{equation}
|\psi_{\alpha}\rangle=\frac{1}{\sqrt{2}}(|\!\uparrow\downarrow\rangle + e^{i\alpha}|\!\downarrow\uparrow\rangle), \quad
|\phi_{\alpha}\rangle=\frac{1}{\sqrt{2}}(|\!\uparrow\uparrow\rangle + e^{i\alpha}|\!\downarrow\downarrow\rangle).
\end{equation}
(In our case the Landau level index $i=1,2$ replaces the spin index $\uparrow$, $\downarrow$.)
Then there is finite entanglement and finite violation of the Bell inequality for arbitrarily strong dephasing \cite{Sam03}, and
moreover there is the same one-to-one relation between degree of entanglement and violation of the Bell inequality as for
pure states. All this no longer holds for non-zero mixing ($\xi \neq 0$), when the maximally entangled electron-hole pair is in
a superposition of $|\phi_{\alpha}\rangle$ and $|\psi_{\alpha'}\rangle$. Then the entanglement disappears for a finite 
dephasing strength and the Bell inequality is no longer capable of unambiguously detecting entanglement.

\section*{Acknowledgements}
This work was supported by the Dutch Science Foundation NWO/FOM and by the U.S. Army Research Office (Grant No. DAAD 19-02-0086).

\end{document}